\documentstyle[11pt,moriond,epsfig]{article}

\def\ra{\rightarrow}

\def\be{\begin{equation}}
\def\ee{\end{equation}}
\def\bea{\begin{eqnarray}}
\def\eea{\end{eqnarray}}

\newcommand {\fpair}            {\mbox{${\mathrm f}\overline{\mathrm f}$}}

\newcommand {\ipb}              {\mbox{pb$^{-1}$}}

\newcommand {\mee}               {{\mathrm e}^+ {\mathrm e}^-}

\newcommand {\sele}             {\tilde{\mathrm e}}

\newcommand {\stau}             {\tilde{\tau}}

\newcommand {\nt}               {\tilde{\chi}^0}

\newcommand {\neutralino}    {\tilde{\chi }^{0}_{1}}

\newcommand {\WW}            {{\mathrm W}^+{\mathrm W}^-}

\newcommand {\roots}            {\sqrt{s}}
\newcommand {\Rparity}          {$R$-parity}

\newcommand{\lb}             {$\lambda$}
\newcommand{\lbp}            {$\lambda^{'}$}

\begin{document}


\vspace*{4cm}
\title{OTHER SEARCHES AT LEP}

\author{ S. BRAIBANT
}

\address{CERN, EP Division, CH-1211 Geneva 23, Switzerland}

\maketitle\abstracts{
During the year 2000, LEP has been operated at centre-of-mass 
energies up to $\roots =$ 209~GeV.
New particle searches have been performed using these data samples.  
Model independent limits on the production cross-sections and
mass limits in the context of the Minimal Supersymmetric 
Standard Model (MSSM) assuming \Rparity\ violation  and 
in the context of gauge-mediated
supersymmetry breaking theories (GMSB) are presented. Searches for technicolor,
excited leptons and leptoquarks are also reviewed. 
}

\section{Introduction}

In the year 2000, the LEP $\mee$ collider has been operated 
at centre-of-mass energies up to 209~GeV. 
In this year, each of the four LEP experiments, 
ALEPH, DELPHI, L3 and OPAL, has collected
an integrated luminosity of about 220~$\ipb$. Using all data
collected between 1995-2000, the total luminosity per experiment is about 
650--700~$\ipb$.

With these data samples, 
searches for \Rparity\ violating decays of 
supersymmetric (SUSY) particles or searches in the context of gauge-mediated
supersymmetry breaking theories are performed and are here described.  
Searches for  technicolor,
excited leptons and leptoquarks are also briefly reviewed. 
A full description 
of the results presented here can be found in~\cite{ref:ALL}.
All results described here are 
preliminary and constitute a representative selection of results from 
each collaboration. 
In most of the searches, some candidates are selected
by the analyses but their number is compatible with the expected
background from Standard Model (SM) processes.
Therefore, 95\% confidence level (C.L.) upper limits 
on the production cross-section are computed and mass limits are derived.

\section{Searches for \Rparity\ Violation Decays of Supersymmetric Particles}

In this section,  a few examples of searches for 
\Rparity\ ($R = -1^{(2S+3B+L)}$) violating decays 
of supersymmetric particles are presented. 
When \Rparity\ is not conserved, the lightest SUSY particle (LSP) 
will decay and the topologies differ significantly from the
ones with conserved \Rparity.

With the MSSM particle content,  \Rparity\ violating interactions are 
parametrised with a gauge-invariant super-potential that includes the 
following Yukawa coupling terms: 
\bea
W_{RPV}  = 
    \lambda_{ijk}      L_i L_j {\overline E}_k
 +  \lambda^{'}_{ijk}  L_i Q_j {\overline D}_k
 +  \lambda^{''}_{ijk} {\overline U}_i {\overline D}_j {\overline D}_k, 
\label{lagrangian}
\eea
where $i,j,k$ are the generation indices of the super-fields 
$L, Q,E,D$ and $U$. $L$ and $Q$ are lepton and quark left-handed doublets,  
respectively. 
$\overline E$, $\overline D$ and $\overline U$ are right-handed 
singlet charge-conjugate super-fields for the charged 
leptons and down- and up-type quarks, respectively.
This makes a total of 45 parameters in addition to those 
of the \Rparity\ conserving MSSM.

Pair-production of gauginos (charginos and neutralinos) and 
sfermions (scalar fermions) is usually assumed but searches for single 
resonant production via the exchange of a sneutrino are also performed.
Two different scenarios are probed. In the first scenario, the decays 
of sparticles via the lightest neutralino, $\nt_1$, are considered, 
where $\nt_1$ is treated as the LSP and assumed to 
decay via \Rparity\ violation.
These are denoted as ``indirect decays''. In the second case,
``direct'' decays
of sparticles to SM particles are investigated.  
In this case, the sparticle considered is assumed
to be the LSP, such that \Rparity\ conserving decay modes do not
contribute.

\begin{figure}[t] 
\centering
\begin{tabular}{cc}
\epsfig{file=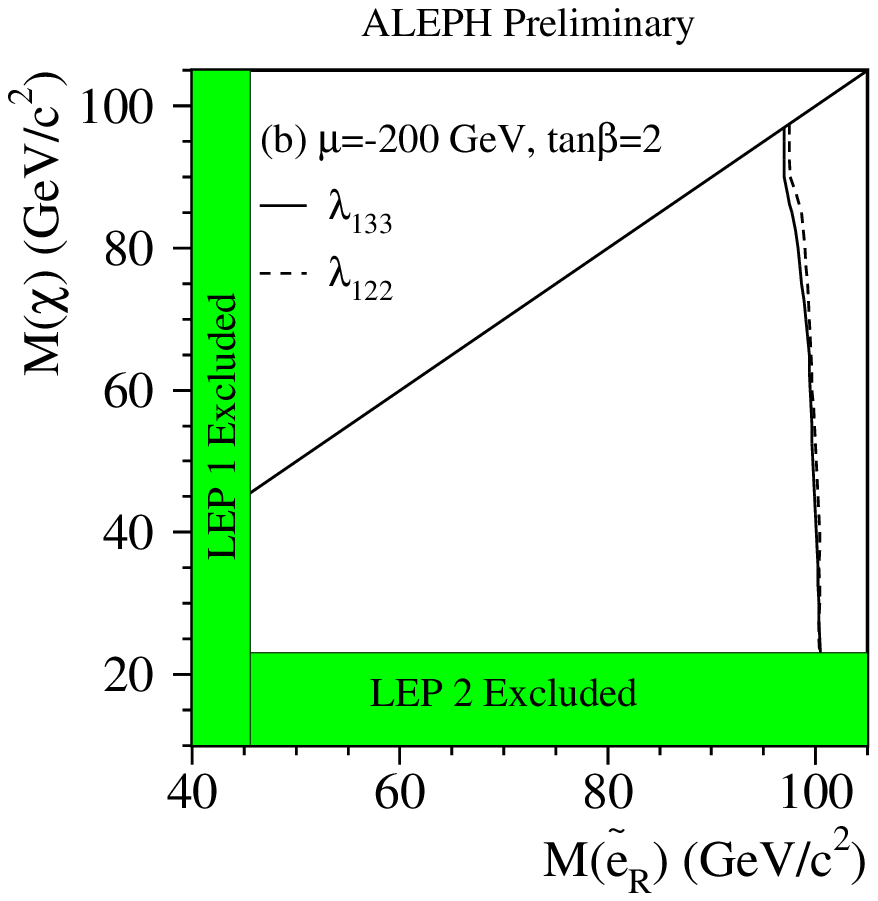,height=5.3cm} 
 & \epsfig{file=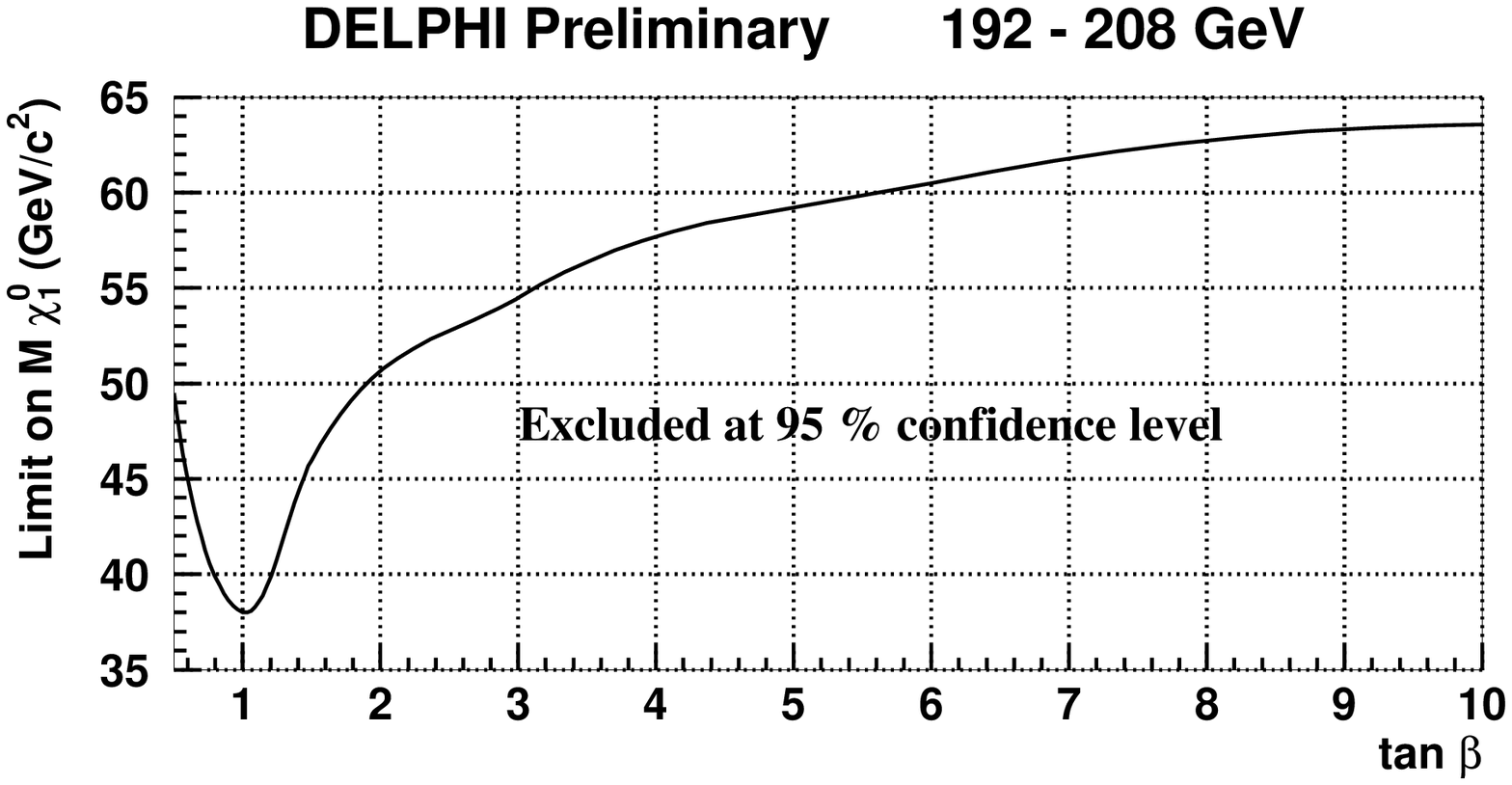,height=5.3cm} \\ 
\end{tabular}
\vspace{10pt}
\caption{
Left plot: 95\% C.L. lower mass limits for a right-handed selectron 
using tan $\beta =$ 2 and $\mu = - 200$~GeV from the ALEPH Collaboration. 
The exclusions are shown for 
\lb$_{133}$ (solid line) and for \lb$_{122}$ (dashed line).
Right plot: 95\% C.L. limits on the $\neutralino$ (LSP) mass as a function of
$\tan \beta$ using $m_0=500$~GeV from the DELPHI Collaboration.}
\label{fig:rpvfig}
\end{figure}

As an example, Fig.~\ref{fig:rpvfig} (left plot)  shows the 95\% C.L. 
lower mass limits for right-handed selectron using a \lb\ coupling, 
while the right plot shows the 95\% C.L. limits for a \lbp\ coupling 
on the $\neutralino$ (LSP) mass as a function of
$\tan \beta$ using $m_0=500$~GeV.
A complete set of results regarding all \lb-like couplings 
for gauginos and sfermions can be found in~\cite{ref:ALL}. In all cases, 
no evidence for supersymmetry with \Rparity\ violation is found and limits are
placed on the production cross--sections, on the sparticle masses and on the 
MSSM parameter space.

\section{Searches for Gauge Mediated Supersymmetry Breaking Signatures}
\begin{figure}[t] 
\centering
\begin{tabular}{cc}
\epsfig{file=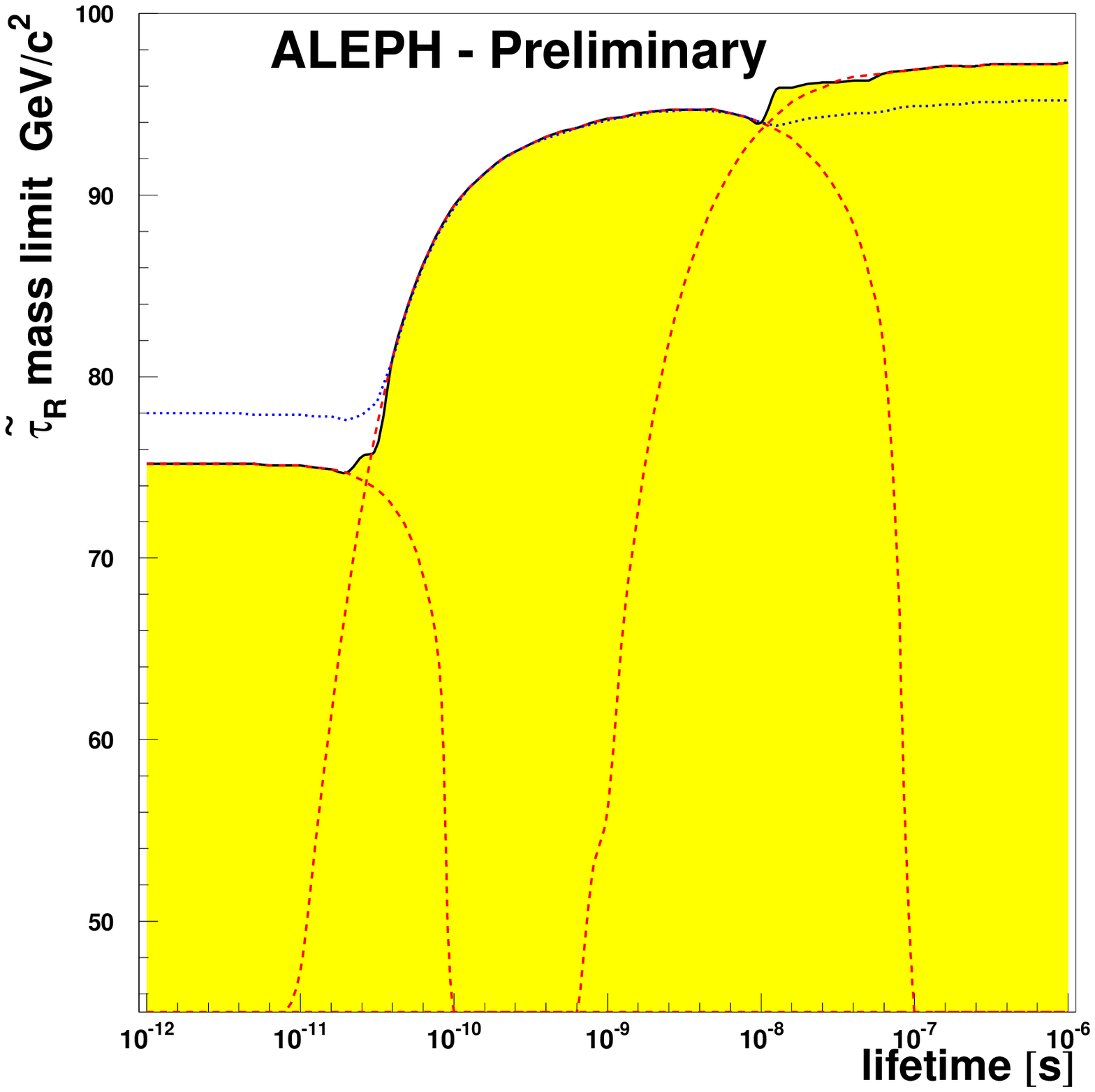,height=6cm} 
 & \epsfig{file=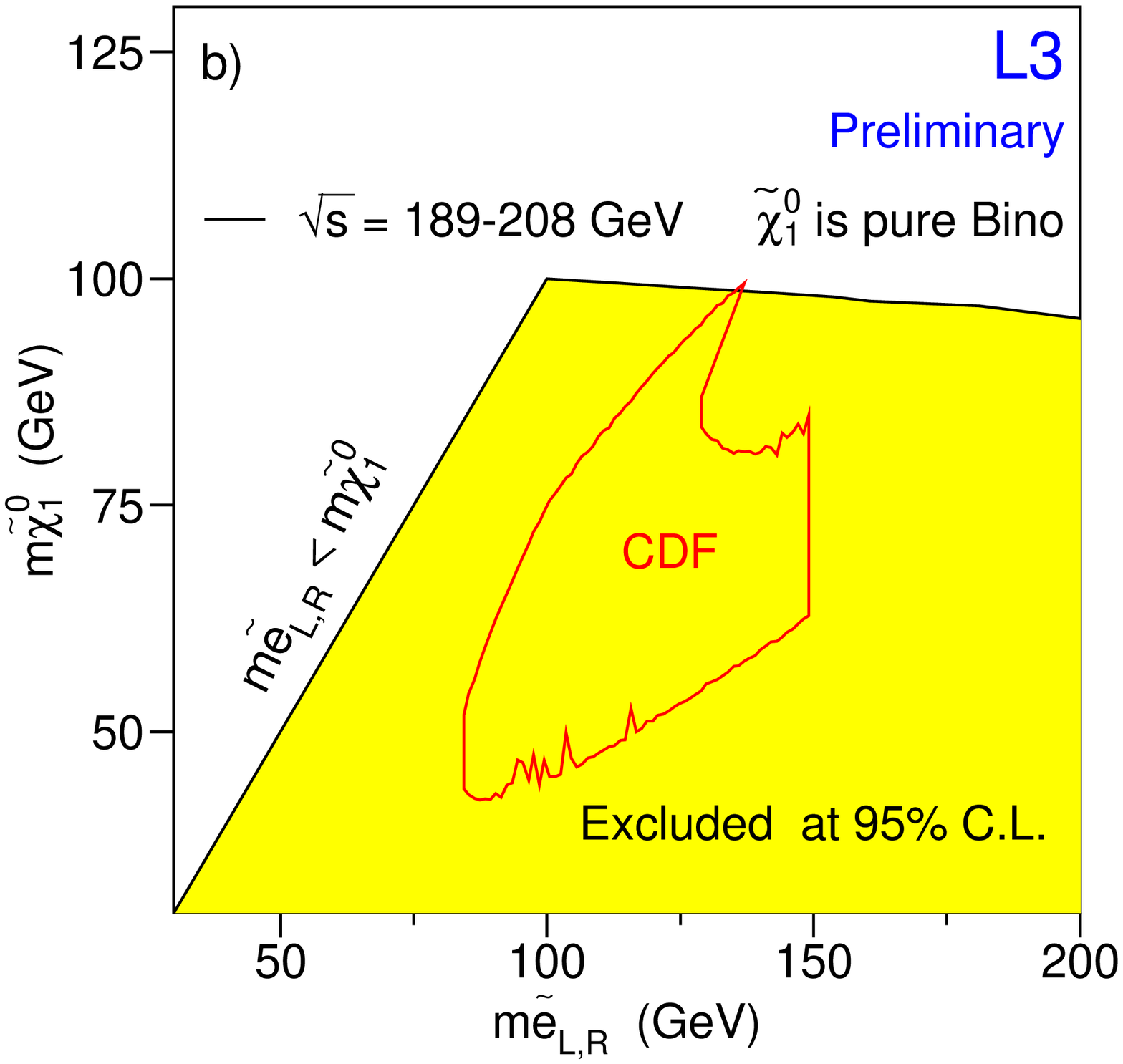,height=6cm} \\ 
\end{tabular}
\vspace{10pt}
\caption{
Left plot: 95\% C.L. excluded mass regions for right-handed staus (NLSP) as a 
function of the NLSP lifetime from the ALEPH Collaboration and combining three
different searches.  
Right plot: 95\% C.L. excluded region in 
the ($m_{\sele_R}$, $m_{\neutralino}$) 
mass plane assuming 
a $\neutralino$ NLSP from the L3 Collaboration; this limit is 
derived from the search for acoplanar photons assuming a zero lifetime NLSP.}
\label{fig:gmsb}
\end{figure}

In this section, searches in the context of a
model called Gauge Mediated Supersymmetry Breaking (GMSB) are presented. 
In this model, supersymmetry is broken via the usual 
gauge interactions in a hidden sector, which 
couples to the visible sector of the SM and SUSY 
particles via a messenger sector. 
In this GMSB model, the supersymmetric partner of the graviton, the gravitino 
$\tilde{G}$ is assumed to be the LSP and the next-to-LSP (NLSP) can either 
be the lightest neutralino, $\tilde{\chi}_1^0$, 
or a right-handed slepton, $\tilde{l}_R$. 
The NLSP decay length is unconstrained and 
all possible decay lengths between zero and infinity have to be considered,
suggesting to explore many different final state topologies.
No significant excesses with respect to SM background 
are observed. Therefore,  model independent limits on the production 
cross--sections and mass limits within the context of GMSB are derived. 
As an example, Fig.~\ref{fig:gmsb} (left plot)  shows the 95\% C.L.
excluded $\stau_R$ mass (GeV) as a function of the lifetime (s), assuming a 
$\stau_R$ NLSP and combining searches exploring various regions of the NLSP 
lifetime. 
The very short lifetime range is 
covered by the search for events with a pair of acoplanar leptons; 
the intermediate lifetime
range is covered by the searches for events containing tracks 
with large impact parameters and kinks; 
and the long lifetime range is covered by the search for heavy 
stable charged particles. 
Fig.~\ref{fig:gmsb} (right plot) shows the 
95\% C.L. excluded region in the ($\sele_R$, $\neutralino$) 
mass plane assuming 
a $\neutralino$ NLSP; this limit is 
derived from the search for events with acoplanar 
photons and assuming a zero lifetime NLSP.

\section{Technicolor, Excited Leptons and Leptoquarks}
\begin{figure}[t] 
\centering
\begin{tabular}{cc}
\epsfig{file=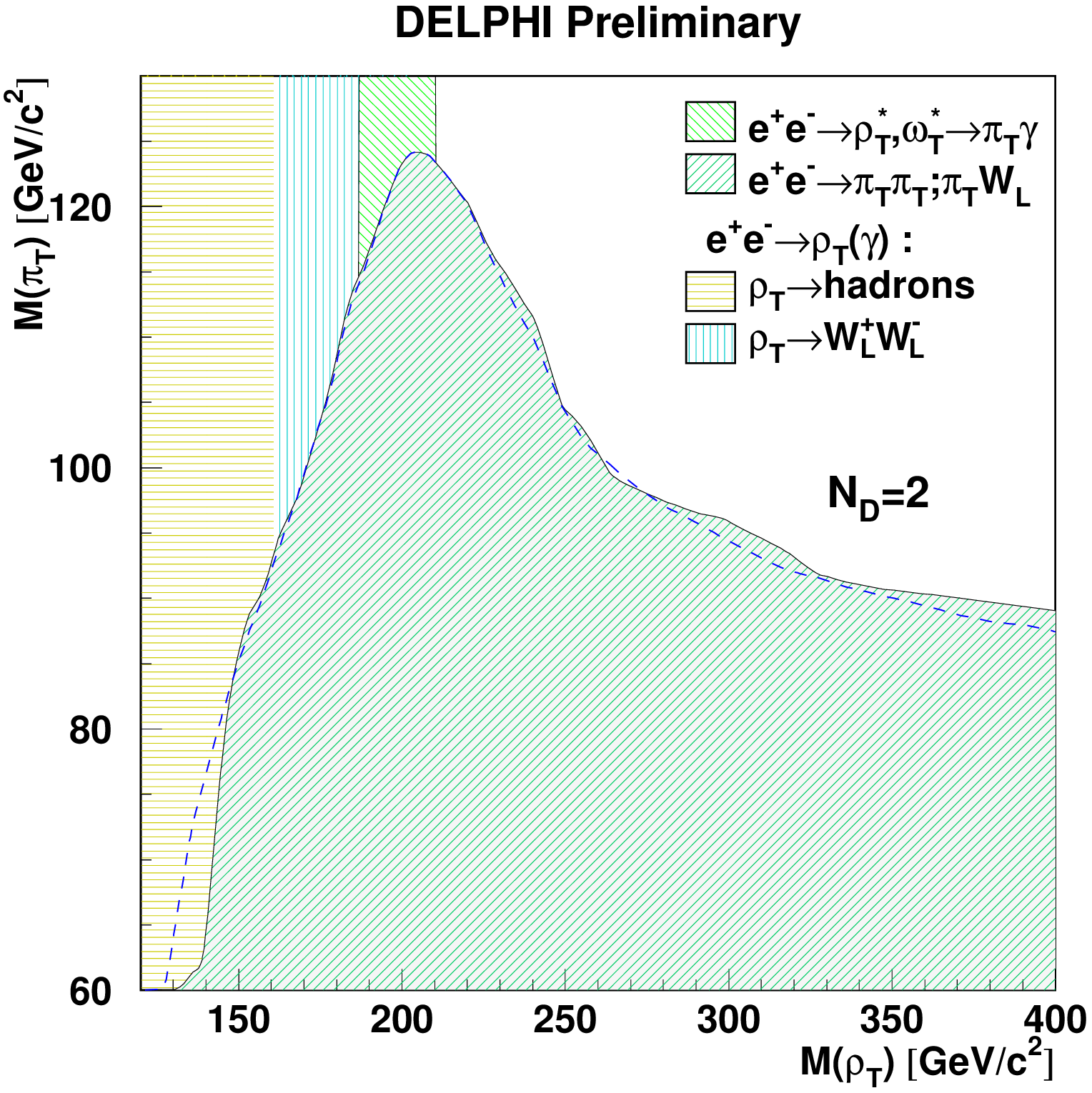,height=6.5cm} 
 & \epsfig{file=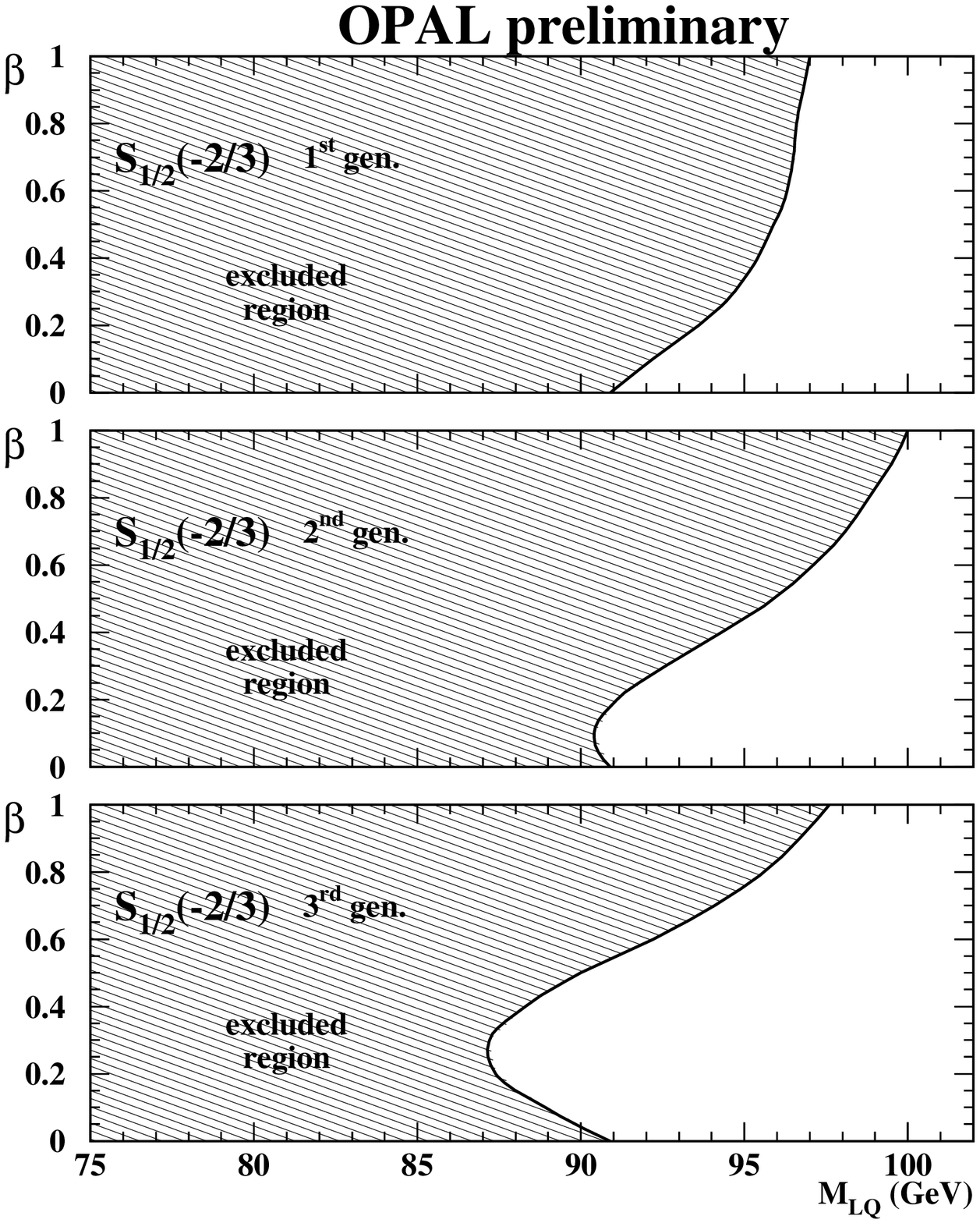,height=6.5cm} \\ 
\end{tabular}
\vspace{10pt}
\caption{
Left plot: 95\% C.L. excluded mass regions 
in the ($m_{\pi_T}$, $m_{\rho_T}$) from the DELPHI Collaboration.
Right plot: regions of the plane 
$\beta$~{\it vs}~M$_{\mathrm{LQ}}$
excluded at 95$\%$ C.L. for the scalar state 
${S}_{1/2}(-2/3)$, with possible values of $\beta$ in the range between
0 and 1 from the OPAL Collaboration.    } 
\label{fig:others}
\end{figure}

Technicolor is a viable alternative to the Higgs mechanisms for 
generating gauge boson masses. DELPHI has searched for many different channels
such as $\mee \ra  \rho_T \ra \fpair$ (below the $\WW$ threshold),
$\mee \ra \rho_T \ra \WW $ (above the $\WW$ threshold) or 
$\mee \ra \rho_T \ra \pi_T^+ \pi_T^-$, followed by 
$\pi_T^+ \ra  \overline{b} c, \overline{b}u$. Good agreement 
is observed with the SM expectation in all channels studied.
This is translated into an excluded region at the 95\% C.L. level 
in the ($m_{\pi_T}$, $m_{\rho_T}$) 
plane as shown in the left plot of Fig.~\ref{fig:others}.     
The $\rho_T$ production is excluded for $m_{\rho_T} < $~202~GeV.

Another interesting topic consists in the search for charged excited leptons
with photonic decays as performed by the OPAL Collaboration.  
In the pair production channel,
$\ell^{*+} \ell^{*-} \rightarrow \ell^+ \ell^- \gamma \gamma$,
no excess of events above the expected background is observed
and limits on the mass of the excited leptons are inferred
with the coupling assumption $f=f^\prime$ (for
the photonic branching ratio), as
shown in Table~\ref{tab:mass_limits}.

\begin{table}
\caption{\label{tab:mass_limits}
  95\% confidence level lower mass limits on excited
leptons extracted from pair production searches performed by the OPAL 
Collaboration.  The limits are calculated assuming $f=f^\prime$.}
\begin{center}
\begin{tabular}{|c||c|} \hline
Flavour & Mass limit (GeV) \\ \hline\hline
$\rm e^*$ & 102.9 \\ \hline 
$\mu^*$   & 102.9 \\ \hline
$\tau^*$  & 102.8 \\ \hline
\end{tabular}
\end{center}
\end{table}

A search for events with pair production of leptoquarks is also performed.
The search is done assuming that leptoquarks are produced via
couplings to the photon and the Z$^{0}$ and promptly decay 
into a lepton and quark. Moreover the simplifying assumption is made
that only couplings within a single generation of leptons exist.
This leads to final states characterized by high multiplicities
and by the presence of isolated and energetic charged leptons
or missing energy if neutrinos are present.
No significant evidence for leptoquark pair production is observed 
in the data and
lower limits on the leptoquark masses are derived 
as a function of the branching ratio, $\beta$, of
decay into a charged lepton and a quark. As an example, Fig.~\ref{fig:others} 
(right plot) shows the regions of the plane 
$\beta$~{\it vs}~M$_{\mathrm{LQ}}$
excluded at 95\% C.L. for the scalar state 
${S}_{1/2}(-2/3)$,
with possible values of $\beta$ in the range between
0 and 1.

\section{Conclusions}
LEP has been a great success until its very end allowing 
a multitude of searches for new particles. These searches have  
been performed using a total integrated luminosity of about 650--700~$\ipb$, 
collected by experiment, at centre-of-mass energies up to 209~GeV. 
A number of preliminary updates of these searches for new particles have 
been performed and interpreted in various models. 
No significant evidence for new physics is observed.

\end{document}